\documentclass[aps,pra,reprint,twocolumn,superscriptaddress,nofootinbib]{revtex4-2}

\usepackage{amsmath}
\usepackage{amssymb}
\usepackage{graphicx}
\usepackage{bm}
\usepackage{hyperref}
\usepackage{float}

\begin{document}

\title{Radiation-pressure induced chaos in multipendular Fabry--Perot resonators}

\author{V. Pierro}
\affiliation{DIIMA, Faculty of Engineering, University of Salerno, Via Ponte Don Melillo, 84084 Fisciano (SA), Italy}

\author{I. M. Pinto}
\affiliation{DIIMA, Faculty of Engineering, University of Salerno, Via Ponte Don Melillo, 84084 Fisciano (SA), Italy}

\date{Received 28 October 1993; accepted for publication 7 December 1993. Communicated by J.P. Vigier}

\begin{abstract}
The occurrence of chaos in multipendular FP resonators (without light propagation delay) is illustrated. The probability of evolving a chaotic solution in a given time, starting from initial conditions contained in a neighbourhood of a stable equilibrium point, is estimated, and its dependence on the most relevant system parameters is investigated. The possible relevance of these results for interferometric gravitational wave antennas is discussed.
\end{abstract}

\maketitle

Radiation-pressure induced multistability in pendular Fabry--Perot resonators (FPRs) was predicted and observed by Dorsel et al.~\cite{ref1,ref2}, and indicated as a potentially effective tool for precise mirror confinement in two and three mirror schemes by Meystre et al.~\cite{ref3,ref4,ref5}.

The possible occurrence of chaos in long single-pendular FPRs, resulting from the combined effect of radiation pressure and light propagation delay described by a nonlinear delay-differential equation, and its potential relevance for interferometric gravitational wave antennas (GWAs) was first suggested by Deruelle, Tourrenc and co-workers~\cite{ref6,ref7,ref8,ref9,ref10}.

This last issue was further investigated by Meers and MacDonald~\cite{ref11}, who presented a linear stability analysis of a complete interferometer (with or without light recycling) and by Solimeno et al., who discussed torsional (radiation-pressure driven) instability in freely swinging possibly misaligned pendular FPRs~\cite{ref12}.

In this Letter we re-examine the whole question in a nonlinear dynamical system perspective, starting from noting that interferometric GWAs use multipendular FPR arms for efficient mechanical insulation from ground seismic noise~\cite{ref13}. FPR dynamics take therefore place in $2N$-dimensional phase space ($N>1$) and thus can be chaotic, even if light propagation delay effects can be 
neglected\footnote{As is well known, the dynamics of a one degree of freedom autonomous Hamiltonian system without delay cannot be chaotic~\cite{ref14}.}.

In the following we refer to the simplest model sketched in Fig.~1, a plane double-pendular FPR, and use the following dimensionless variables and parameters:
\begin{equation}
\xi_i = \frac{x_i}{\lambda}, \quad \bar{t} = \left(\frac{g}{L_2}\right)^{1/2} t,
\end{equation}
\begin{equation}
\Lambda = \frac{L_2}{L_1}, \quad \mu = \frac{M_2}{M_1 + M_2}, \quad \Pi = \frac{2P}{c M_2 g},
\end{equation}
where $g$ is the terrestrial gravity acceleration, $c$ the velocity of light in vacuo, $\lambda$ the light wavelength, and $P$ the laser light power.

\begin{figure}[htbp]
\centering
\includegraphics[width=\columnwidth]{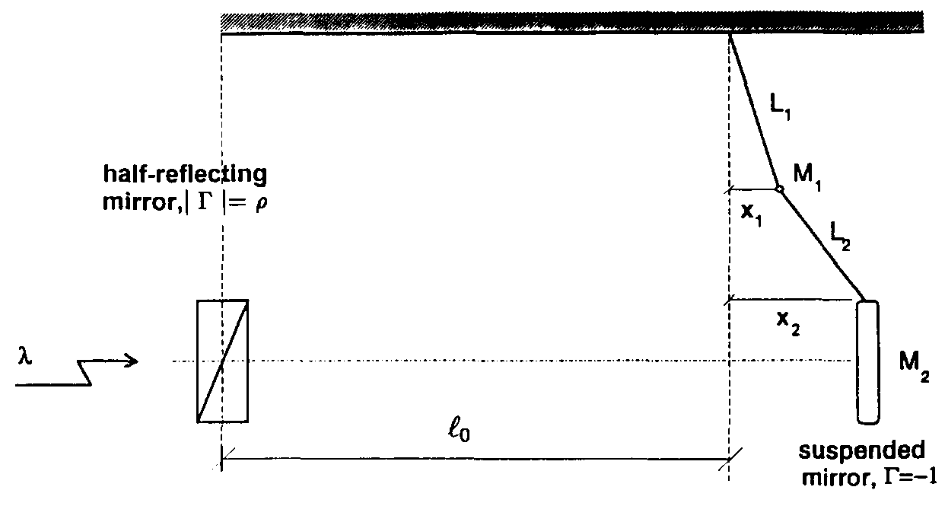}
\caption{Sketch of plane double pendular FPR.}
\label{fig:fig1}
\end{figure}

\begin{table}[htbp]
\caption{Test-case parameter values}
\label{tab:tab1}
\centering
\renewcommand{\arraystretch}{1.35}
\begin{tabular*}{\columnwidth}{@{\extracolsep{\fill}}cc}
\hline\hline
Parameter & Value \\
\hline
$\mu$ & 0.5 \\
$\Lambda$ & 1.0 \\
$\rho$ & 0.995 \\
$\phi$ & 0.0 \\
$\lambda$ & 514 nm \\
$M_2$ & 400 kg \\
\hline\hline
\end{tabular*}
\end{table}

Light propagation delays should be taken into account whenever the cavity storage time is comparable to the pendular period(s), viz.
\begin{equation}
\Omega \frac{\mathcal{F}\ell}{\pi c} \sim 1,
\end{equation}
$\Omega$ being a typical pendular eigenfrequency, $\mathcal{F}$ the cavity finesse and $\ell$ the cavity length. For quite realistic system parameters like, e.g., those collected in table~1, the l.h.s. of (3) is $\approx 10^{-2}$ and light propagation delays can be neglected. The dynamics can then be described, on timescales much shorter than the pendular damping times, by the following Hamiltonian\footnote{Energy scaling to $M_2 g \lambda^2 / L_2$ is implied.},
\begin{align}
\mathcal{H} &= \frac{1}{2}\left( \frac{\mu}{1 - \mu} \eta_1^2 + \eta_2^2 \right) + \frac{1}{2}\frac{\mu + \Lambda}{\mu}\xi_1^2 - \xi_1 \xi_2 \nonumber \\
&\quad + \frac{1}{2}\xi_2^2 + \mathcal{U}_{\text{rad}}(\xi_2),
\end{align}
where
\begin{equation}
\eta_1 = \frac{1 - \mu}{\mu}\dot{\xi}_1, \quad \eta_2 = \dot{\xi}_2
\end{equation}
are the canonical momenta conjugate to $\xi_1, \xi_2$, the dot representing differentiation with respect to $\bar{t}$,
\begin{equation}
\mathcal{U}_{\text{rad}}(\xi_2) = \frac{L_2}{\lambda}\,\frac{\Pi}{2\pi} \arctan\left( \frac{1 - \rho}{1 + \rho} \cot(2\pi\xi_2 + \tfrac{1}{2}\phi) \right)
\end{equation}
is the potential of the radiation force, $\rho$ is the modulus of the fixed-mirror reflection coefficient, and $\phi$ the detuning parameter~\cite{ref6},
\begin{equation}
\phi = \frac{4\pi}{\lambda}(\ell_0 - \ell_k) \in [-\pi, \pi],
\end{equation}
$\ell_0$ being the cavity rest length, and $\ell_k$ the resonant cavity length closest to $\ell_0$ (see Fig.~1).

Initial conditions will be assumed to lie in a tiny neighbourhood of the lowest-energy equilibrium point (see appendix~A), as a result of cold-damping. Electromagnetic cold-dampers of the pendular motion are used in interferometric GWAs~\cite{ref13} both to prevent large seismic noise-driven mirror oscillations at the pendulum resonances, and to prepare the system close to its working position before the laser is switched on. Cold damping of the pendular modes to within absolute displacements $\sim 10^{-6}\text{ m}$ has been reported~\cite{ref15}, and could be likely reduced by a factor $\sim 10^2$ using improved PSD designs~\cite{ref16}. Hence we take
\begin{equation}
\{ |\xi_i - \xi_i^{(\text{eq})}| \le 0.005, \; |\eta_i| \le 0.005, \; i=1, 2 \},
\end{equation}
to represent the set of possible initial conditions.

Extensive numerical simulations including phase-space trajectories, Poincar\'e sections, power spectra, and Lyapunov exponents provided strong numerical evidence that the above system is non-integrable and exhibits chaotic dynamics~\cite{ref17}.

In the following we shall report a number of illustrative results for the test-case referred in Table~1.

In Figs.~2--4 the principal Poincar\'e sections of the (nonlinear) oscillations are displayed for values of $\mathcal{H}$ exceeding the equilibrium-point energy by $\Delta \mathcal{H} = 1.0\times 10^{-5}, 1.38\times 10^{-5}, 2.0\times 10^{-5}$, in scaled units\footnote{These values are consistent with (8).}. The coordinates used in Figs.~2--4 are defined in appendix~B. The abrupt onset and spreading of connected chaos is clearly exhibited. As a further illustration, typical computed frequency spectra of regular and stochastic motions are shown in Fig.~5.

\begin{figure}[htbp]
\centering
\includegraphics[width=\columnwidth]{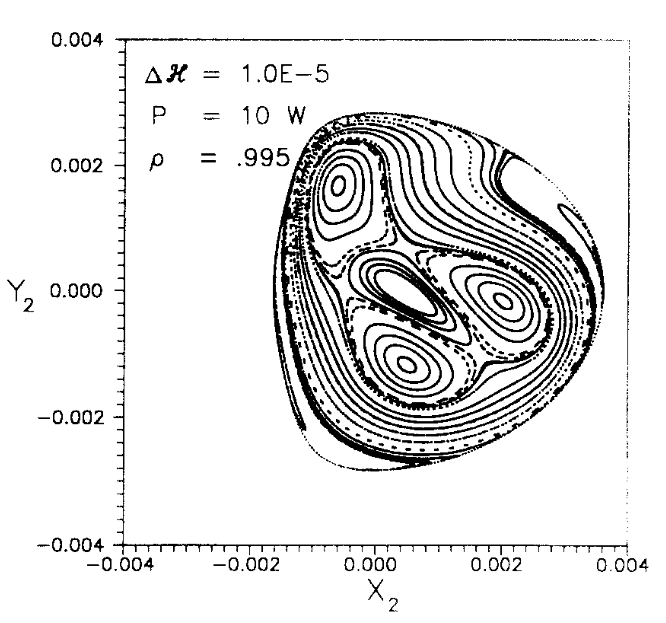}
\caption{Poincar\'e section of phase flow.}
\label{fig:fig2}
\end{figure}

\begin{figure}[htbp]
\centering
\includegraphics[width=\columnwidth]{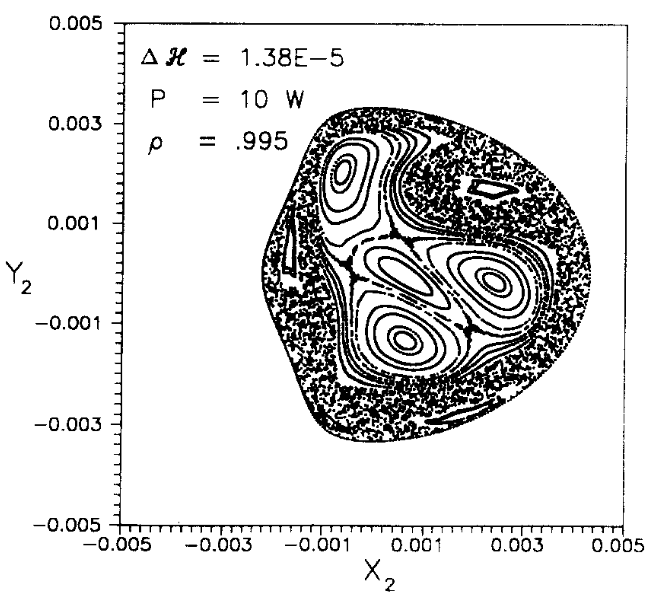}
\caption{Poincar\'e section of phase flow.}
\label{fig:fig3}
\end{figure}

\begin{figure}[htbp]
\centering
\includegraphics[width=\columnwidth]{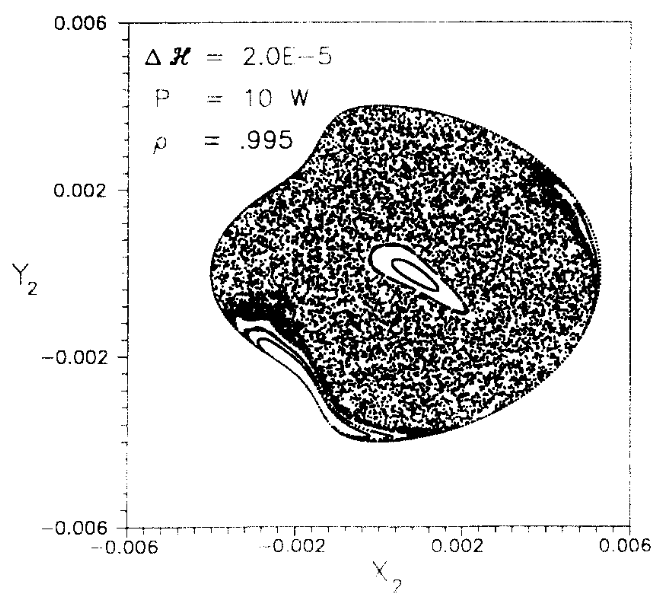}
\caption{Poincar\'e section of phase flow.}
\label{fig:fig4}
\end{figure}

The next natural question concerns the probability of developing chaos within a typical observation time. An illustrative answer has been obtained by evolving the system until $\bar{t} = \bar{t}_{\text{max}} = 10^3$, starting from initial conditions arranged in a regular square lattice of $10^4$ points in the phase-space cell~(8), and computing for each trajectory the Sinai entropy $S$ (sum of the positive Lyapunov exponents). The statistical density of $S$ is shown in Fig.~6, for $P=25\text{ W}$. The reciprocal of $S$ provides an estimate of the time scale over which information about the state of the system is lost (Pesin theorem~\cite{ref14}). Initial conditions for which $S^{-1} \ge \bar{t}_{\text{max}}$ are consistently classified as yielding regular motions. The fraction of initial conditions originating chaotic trajectories provides an estimate of the overall probability of chaos occurrence, and is found to depend on the main system parameters $\rho$ and $\Pi$ as shown in Fig.~7.

It is seen that chaos is likely to occur within $0 \le \bar{t} \le 1000$ for all values of $\rho$ and $\Pi$ of practical interest.

It should be recognized that the model considered above is fairly simplified. In this connection the following comments are in order: (i) on the timescales where chaos has been found to blow up, the friction-free (Hamiltonian) model appears justified, in view of the comparatively large envisaged pendular damping times ($\sim Q \sim 10^6$, in scaled units~\cite{ref13}); (ii) addition of more pendular sections and/or further mechanical (e.g., tilting and swiveling) or electromagnetic (Gauss--Hermite modes~\cite{ref18}) degrees of freedom would essentially increase the phase-space dimension, and likewise make the dynamics even more complicated~\cite{ref19}, making chaotic diffusion (the Arnold web) also possible; (iii) Hamiltonian chaos is conjectured to be generally robust with respect to the addition of (weak, classical) noise~\cite{ref14}. This gives some confidence that including noise (seismic, Brownian, laser) should not alter significantly the picture.

\begin{figure*}[t]
\centering
\includegraphics[width=2\columnwidth]{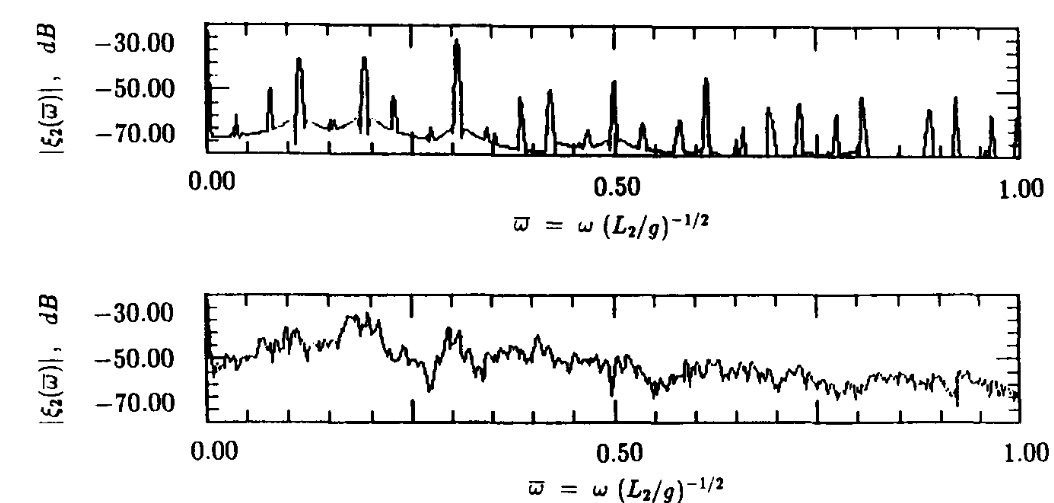}
\caption{Frequency spectra of prototype regular and chaotic motions. Top: regular motion, initial conditions $\xi_1 = 4.8122\times 10^{-4}, \xi_2 = 1.4437\times 10^{-3}, \eta_1 = -4.3254\times 10^{-3}, \eta_2 = 2.9817\times 10^{-3}$. Bottom: chaotic motion, initial conditions $\xi_1 = 3.5791\times 10^{-3}, \xi_2 = 5.9375\times 10^{-3}, \eta_1 = -4.1191\times 10^{-8}, \eta_2 = 2.8395\times 10^{-8}$. Both cases: $\Delta\mathcal{H} = 1.38\times 10^{-5}$, FFT size 1024, 10 averages.}
\label{fig:fig5}
\end{figure*}

\begin{figure}[htbp]
\centering
\includegraphics[width=\columnwidth]{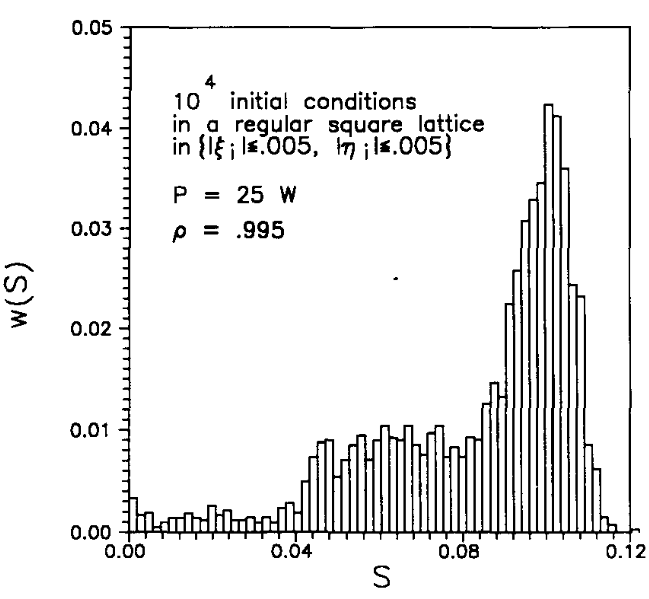}
\caption{Statistical density of the Sinai entropy.}
\label{fig:fig6}
\end{figure}

\begin{figure}[htbp]
\centering
\includegraphics[width=\columnwidth]{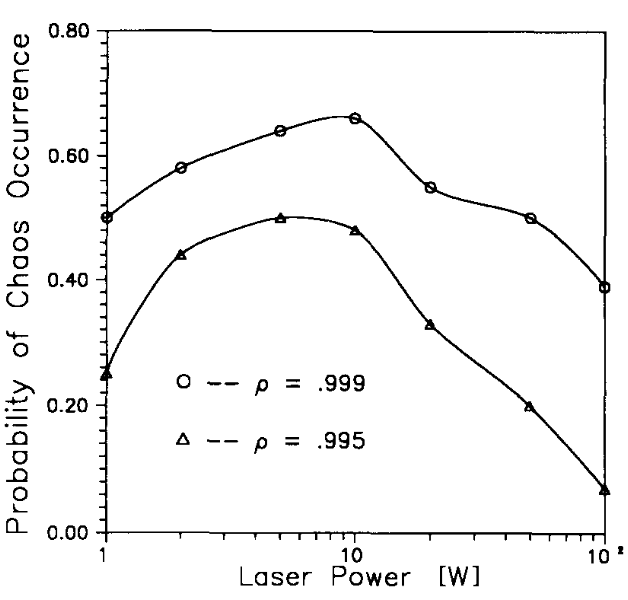}
\caption{Probability of chaos occurrence within $\bar{t} \le 1000$ versus $P$.}
\label{fig:fig7}
\end{figure}

The possible relevance of these results as well as those in refs.~\cite{ref6,ref7,ref8,ref9,ref10} to the operation of very long baseline interferometric (VLBI) GWAs with FPR arms is doubtful, however.

As a matter of fact, experimental interferometer noise characteristics agree quite well with theoretical predictions based on standard linear theory, except perhaps for the wideband excess noise usually observed in the frequency band between the seismic and shot noise lines, well above the Brownian noise level. This spectral range is of potential interest to pulsar GW astronomy, and happens to overlap with the typical computed spectrum of chaotic noise (see Fig.~5). It is tempting to speculate that chaos could be responsible for this disagreement.

In real (or foreseen) GW interferometric antennas FPR cavities are embedded in a sophisticated feedback control system~\cite{ref13}. In the simplest case, the latter includes, besides the already mentioned cold dampers of the pendular modes\footnote{Computed chaotic vibrations show typical amplitudes $10^{-12}\sim 10^{-10}\text{ m}$ (see Figs.~2--5) well below the cold damping domain of action. Accordingly, cold damping could be only relevant in setting the size of the phase-space cell where the system initial conditions are expected to lie.}, a fringe-locking servo-loop, which continuously adjust the optical cavity length, by both applying suitable electromagnetic forces to the end mirrors (at low frequencies, i.e., below $\sim 10\text{ Hz}$), and introducing suitable phase shifts along the light path using Pockels cells (at high frequencies, i.e., above $\sim 10\text{ Hz}$)\footnote{Further control systems are usually needed to control the mirror misalignments, which are not included in the present bare-bones model.}.

The FPR arms dynamics can be therefore quite different from those of freely swinging radiation-pressure driven pendular FPRs, as discussed here and in refs.~\cite{ref6,ref7,ref8,ref9,ref10}. We stress that splitting the problem, i.e., using these solutions to find the response of the fringe-locking servo-loop to chaos, as if it were an external, superimposed noise would be incorrect in principle, and could be misleading.

It is nonetheless our opinion that these results together with those in refs.~\cite{ref6,ref7,ref8,ref9,ref10} give enough motivation for studying in a full nonlinear-dynamical-system approach the whole interferometer, including both the radiation-force driven FPR arms and the fringe-locking feedback-loop servo.

\appendix

\section{Equilibrium points}
Equilibrium points are stationary points of the potential energy, i.e., from (4)
\begin{equation}
-\xi_1 + \xi_2 + \frac{d\mathcal{U}_{\text{rad}}}{d\xi_2} = 0, \quad \frac{\Lambda + \mu}{\mu}\xi_1 - \xi_2 = 0,
\end{equation}
and are thus found by solving
\begin{equation}
\frac{\Lambda}{\Lambda + \mu}\xi_2 + \frac{d\mathcal{U}_{\text{rad}}}{d\xi_2} = 0,
\end{equation}
and/or equivalently locating the stationary points of the function
\begin{equation}
\frac{1}{2}\frac{\Lambda}{\Lambda + \mu}\xi_2^2 + \mathcal{U}_{\text{rad}}(\xi_2),
\end{equation}
sketched in Fig.~8.

The solutions of (A.2) are graphically represented by the intersections between the straight line $y = (\mu/\Lambda + 1)^{-1}x$ representing the (scaled) weight force, and the periodic spiky function $y = -d\mathcal{U}_{\text{rad}}/dx$ representing the (scaled) radiation force (see Fig.~9).

\begin{figure}[htbp]
\centering
\includegraphics[width=\columnwidth]{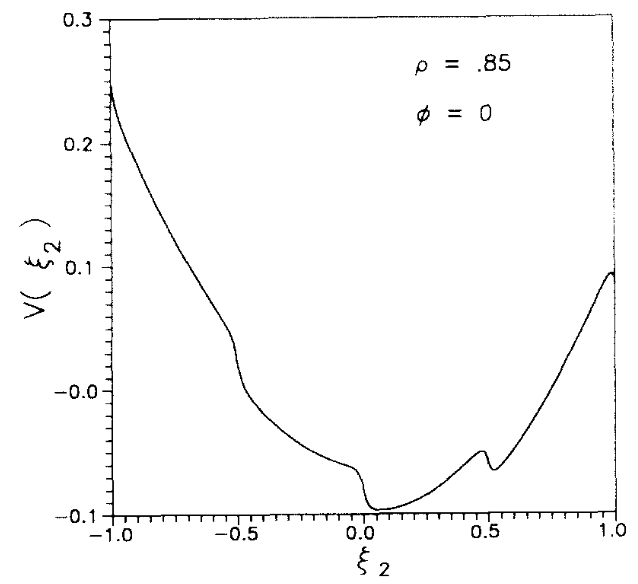}
\caption{Potential function $V(\xi_2)$.}
\label{fig:fig8}
\end{figure}

\begin{figure}[htbp]
\centering
\includegraphics[width=\columnwidth]{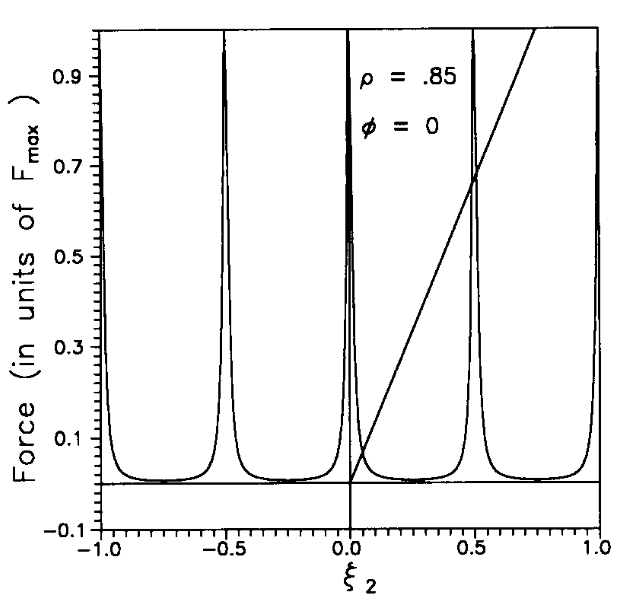}
\caption{The solution of eq.~(A.2).}
\label{fig:fig9}
\end{figure}

The lowest-energy equilibrium point is close to the vertical position $\xi_2 = 0$ (see Fig.~8). Further equilibrium points, if any, occur in pairs, close to the radiation force peaks (see Fig.~9), the point on the right of the peak tip being stable, the other one unstable (see Fig.~8). The number of pairs $N_p$ is
\begin{equation}
N_p = \left[ 2\frac{\Lambda + \mu}{\Lambda} \mathcal{F}_{\text{max}} + \frac{\phi}{4\pi} \right],
\end{equation}
where
\begin{equation}
\mathcal{F}_{\text{max}} = \frac{L_2}{\lambda}\,\Pi \frac{1 + \rho}{1 - \rho}.
\end{equation}
In the very special case where the argument in (A.4) is an integer, the last pair coalesces at the tip of the radiation force peak, and (A.3) exhibits an inflection point here.
\[
\,\,\,
\]
\section{Coordinate transformations}
The coordinates used in Figs.~2--4 are
\begin{equation}
X_i = (\tfrac{1}{2}\Omega_i)^{1/2} q_i, \quad Y_i = (1/2\Omega_i)^{1/2} p_i, \quad i=1, 2,
\end{equation}
where $q_i, p_i$ are the normal coordinates and canonical conjugate momenta of the problem obtained by linearizing (4) in the neighbourhood of a stable equilibrium point $\xi^{(\text{eq})}$ (see appendix~A),
\begin{align}
\Omega_{1,2}^2 &= \frac{1}{2}\Biggl\{ \frac{1+\Lambda}{1-\mu} + \mathcal{U}_{\text{eq}}^{(2)} \mp \Biggl[ \left( \frac{1+\Lambda}{1-\mu} + \mathcal{U}_{\text{eq}}^{(2)} \right)^2 \nonumber \\
&\quad - 4\left( \frac{\Lambda}{1-\mu} + \frac{\Lambda+\mu}{1-\mu}\mathcal{U}_{\text{eq}}^{(2)} \right) \Biggr]^{1/2} \Biggr\},
\end{align}
are the normal mode frequencies, and
\begin{equation}
\mathcal{U}_{\text{eq}}^{(2)} = \left. \frac{d^2\mathcal{U}_{\text{rad}}}{d\xi_2^2} \right|_{\xi_2 = \xi_2^{(\text{eq})}}.
\end{equation}

The normal coordinates $q_i$ are related to the displacements $\xi_i - \xi_i^{(\text{eq})}$ by the (linear, symplectic) transformation
\begin{equation}
\begin{pmatrix} \xi_1 - \xi_1^{(\text{eq})} \\ \xi_2 - \xi_2^{(\text{eq})} \end{pmatrix} = \begin{pmatrix} \beta_1(1 + \mathcal{U}_{\text{eq}}^{(2)} - \Omega_1^2) & \beta_2(1 + \mathcal{U}_{\text{eq}}^{(2)} - \Omega_2^2) \\ \beta_1 & \beta_2 \end{pmatrix} \begin{pmatrix} q_1 \\ q_2 \end{pmatrix},
\end{equation}
where
\begin{equation}
\beta_i = \left( \frac{1-\mu}{\mu}(1 + \mathcal{U}_{\text{eq}}^{(2)} - \Omega_i^2)^2 + 1 \right)^{-1/2}.
\end{equation}
The full Hamiltonian describing the nonlinear oscillations in the neighbourhood of the equilibrium point $\xi^{(\text{eq})}$ accordingly reads\footnote{The (constant) equilibrium point energy has been dropped.}
\begin{align}
\mathcal{H} &= \frac{1}{2}(p_1^2 + \Omega_1^2 q_1^2 + p_2^2 + \Omega_2^2 q_2^2) \nonumber \\
&\quad + \mathcal{U}_{\text{NL}}(\xi_2^{(\text{eq})} + \beta_1 q_1 + \beta_2 q_2),
\end{align}
where $\mathcal{U}_{\text{NL}}$ is the strictly nonlinear part of the radiation potential (6), at $\xi^{(\text{eq})}$, viz.
\begin{equation}
\mathcal{U}_{\text{NL}}(\xi_2) = \mathcal{U}_{\text{rad}}(\xi_2) - \left. \sum_{n=0}^2 \frac{d^n \mathcal{U}_{\text{rad}}}{d\xi_2^n} \right|_{\xi_2 = \xi_2^{(\text{eq})}} \frac{(\xi_2 - \xi_2^{(\text{eq})})^n}{n!}.
\end{equation}

\end{document}